# Many-Task Computing Tools for Multiscale Modeling

Daniel S. Katz, University of Chicago & Argonne National Laboratory
Matei Ripeanu, University of British Columbia
Michael Wilde, Argonne National Laboratory & University of Chicago

As computers have become more powerful, the simulations and data processing applications that use them have become resource hungry, and, at the same time more complex. The complexity of simulations has increased in terms of the number of dimensions (from 1D to 2D to 3D), the set of equations being simulated (from one equation, to multiple equations in one domain, to multiple equations in multiple domains), the number of time, length, or other scales being studied simultaneously, and the level of interaction between scales. In multiscale modeling, several applications are frequently assembled to perform increasingly complex and integrated multiscale analyses. Increasing the scientific insights and productivity from such multiscale methods often demands increasing the number of individual applications that are assembled into a meta-application. This can be achieved by adding another layer of procedure around the individual applications, as is done in parameter sweep, optimization, and uncertainty quantification methods. Such meta-applications can be considered "many-task computing" (MTC) applications, as they are assembled of a series of tasks, each of which may be as complex as a standalone application or simple as a procedure call.

Along the same lines, David Keyes recently argued that today's computational scientists require increased computing performance for eight reasons: resolution, fidelity, dimension, artificial boundaries, parameter inversion, optimal control, uncertainty quantification, and the statistics of ensembles[1]. The last four of these requirements can be addressed by MTC.

The term MTC first appeared in the literature in 2008, introduced to describe a class of applications that did not fit neatly into the categories of traditional High Performance Computing (HPC) or High Throughput Computing (HTC)[2]. Also in 2008, a series of workshops titled "Many-Task Computing on Grids and Supercomputers" began; this workshop has been held at the SC08 – SC10 conferences and will be held again at SC11.

As with traditional HPC, a defining aspect of MTC is the emphasis on performing a large amount of computation in a timespan of hours or days, in order to provide important results in a timely manner. However, in contrast to traditional HPC applications, which tend to be a single program run simultaneously on many nodes of a single cluster or supercomputer (e.g., using MPI), an MTC application is a set of many interdependent distinct tasks, often viewed as a directed graph of data dependencies. MTC tasks typically receive inputs, process to completion, and produce outputs, much like a function, whereas HPC tasks more typically exchange multiple messages with other tasks during their lifetime. While in many cases the data dependencies will be files that are written to and read from a file system shared between the compute resources, MTC does not exclude applications in which tasks communicate in other manners, including in-memory or message-based parameter and result passing.

For many applications, a graph of distinct tasks is a natural way to both conceptualize the computation and to build the application, particularly if some tasks can be performed by existing standalone programs. Structuring a meta-application in this way also enhances the flexibility of execution techniques. For example, it allows tasks to be run on multiple computing sites simultaneously; it simplifies failure recovery by allowing the application to continue when nodes fail (if tasks write their results to persistent storage as they finish); and it permits the application to be tested and run on varying numbers of nodes without any changes to the code of the meta-application.

We believe that many-task computing will make valuable contributions to multiscale modeling. A multiscale application is formed by integrating multiple simulation applications, each of which is designed to operate at a different scale. Multiscale modeling applications can be viewed across a spectrum of coupling, with *loose coupling* at one end of the spectrum, and *tight coupling* at the other end. Beyond the loose-coupling end of the spectrum is manual coupling, where the application is not yet executed as a single application. Here, after a simulation is run, the user examines the output, and then runs another simulation, repeating and adding simulations until the problem is solved, which might take days or weeks. Loose coupling automates this process, and combines the simulations into a single MTC application, where the simulations at different scales read input files and write output files, and some higher level procedure (e.g., a script) orchestrates the simulations, including mapping (and possibly translating) the outputs of one simulation into the inputs for another. In loose coupling, the simulations typically occur in different memory address spaces, and the coupling is done on the scale of minutes to hours. Tight coupling is at the other end of the spectrum, where the simulations typically occupy the same or proximate memory spaces, run simultaneously, and communicate via messages on a scale of seconds to minutes. The choice whether to adopt a loose- or tight-coupling model for a multiscale modeling application often depends on how frequently the interactions between scales occur, and how much work the simulations do independently.

The University of Chicago and Argonne National Laboratory have developed a "parallel scripting" language called Swift[3,4] for assembling and executing MTC applications. The Swift language provides an implicitly parallel and deterministic programming model, which applies external applications to file collections using a functional style that abstracts and simplifies distributed parallel execution. Expressing the higher-level logic of an application in Swift reduces the complexities of repeated execution of domain-specific application programs on large collections of file-based data. In Swift, file system structures are accessible via language constructs and ordinary application programs can be composed into powerful parallel scripts that can efficiently utilize parallel and distributed resources. Swift provides resource orchestration and data management services as data is passed to, from, and between application invocations.

One application in which Swift is currently is multiscale modeling in geophysics for modeling subsurface flows[5] of compounds in groundwater. This application couples continuum and pore-scale simulations. The continuum simulation runs across the entire domain, and after it is run, a second task looks at its output, determining where to run pore-scale simulations and creating the input files for each. After the pore-scale simulations are run, a task takes the outputs of these simulations and the output of the continuum simulation, and builds an input file for a subsequent continuum model. These simulations and "converter" tasks iterate forward in time. Another application for which Swift is being evaluated is multiscale modeling in biomolecular science. Here, Swift can orchestrate tasks that automate the coarse-graining of molecular dynamics simulations[6,7,8]. The questions to be answered by this application are: How many coarse-grained sites are needed? Which atoms are mapped to which coarse-grained sites? What is the potential energy as a function of coordinates of those coarse-grained sites? In a third application, Swift is being studied for use in multiscale modeling of nuclear fuel simulation, where finite element and phase-field modeling can be used together to understand the heating of fuel rods[9].

We are exploring solutions to extend Swift to cover an increasingly large part of the coupling spectrum. This involves a number of issues related to data passing techniques, including transition from file-based to message-based data transfers. In terms of performance, messages keep data in memory and benefit from the speed of memory-to-memory over memory-to-disk copies. Even when files are stored in memory-based filesystems instead of on disk, using shared files through the POSIX API is less efficient than highly optimized message passing. However, the use of sufficiently persistent files does facilitate easier fault recovery and is compatible with many existing applications. Files also provide better separation between the simulations, whereas messages require both message endpoints to exist simultaneously.

A practical programming and performance issue when designing middleware support for MTC applications, whether loosely- or tightly-coupled, is the issue of coordinating the application components and tasks. In the loosely-coupled case, a driver program or script is needed to coordinate the individual tasks, which are typically instantiated each time they are called (although some MTC implementations optimize this by instantiating applications as more persistent services). If each simulation uses the same set of resources, these resources can be kept fully utilized during the overall application's execution. In the tight case, there is just a single executable that includes all the simulations; each is only instantiated once. Each simulation typically uses an independent and static subset of the resources. If each takes the same amount of time, the resources again can be kept busy during the full application execution.

A number of activities that are intended to extend Swift's applicability across the coupling spectrum are underway. These fall into the areas of execution of tasks and data management. In task execution, there are two activities. One is JETS[10], which allows Swift to handle MPI tasks in addition to single-node tasks. The other is Turbine[11], which distributes not only task execution but also program evaluation across a set of resources that communicate through messages. This overcomes bottlenecks in Swift that limit extreme scalability. Task dispatch in Swift is done centrally, which can be limiting when extremely large numbers of tasks need to be executed at high rates (for example, millions of concurrent tasks launched at rates of many thousands of tasks per second).

In data management, three activities are underway. One is "collective data management" (CDM)[12], which can provide hints to the runtime system from the application developer to improve file movement. CDM allows the developer to identify data movement patterns, so that optimized data transfer techniques such as broadcast, scatter, or gather can be applied at the file level. These can either be generic (such as a gather function) or system-specific (such as hardware-assisted broadcast). The other two are related to intermediate file storage. Both continue to use files for passing information between simulations. One, MosaStore[13], stripes files across RAM disk on compute nodes, forming a single distributed filesystem with a shared namespace. It could be used as a cache between the shared file system, which remains as a backing store, and the local tasks. The other, AME[14], uses memory on the compute nodes as a set of local file systems, uses a distributed hash technique to track where files are located, and copies them to nodes as needed to serve as the inputs of new tasks. While AME limits the size of files that it can store, as each file must fit in the memory of a node, it provides excellent performance for many practical application use cases.

We believe that these enhancements will allow Swift's implicitly parallel programming model to cover an increased range of the spectrum of coupled multiscale applications. By exchanging data via files stored in memory, we move towards message passing. By optimizing file I/O, we improve efficiency. By improving task execution rates, we move from a task model to a function model, and possibly towards an executable assembled from multiple simulation tools. We have already shown that Swift works well for loose coupling. Now, we will see how much of the spectrum it can also cover in the direction of traditionally tightly-coupled approaches.

This work was partially supported by the U.S. Department of Energy under the ASCR X-Stack program (contract DE-SC0005380) and under contract DE-AC02-06CH11357, and by the National Science Foundation under grants OCI-721939 and OCI-0944332. This work is licensed under the Creative Commons Attribution-NonCommercial-ShareAlike 3.0 Unported License. To view a copy of this license, visit http://creativecommons.org/licenses/by-nc-sa/3.0/ or send a letter to Creative Commons, 444 Castro Street, Suite 900, Mountain View, California, 94041, USA.

**References:**
[1] David Keyes. Exaflop/s, seriously!, 2010. Keynote lecture for Pan-American Advanced Studies Institutes Program (PASI), Boston University.
[2] Ioan Raicu, Zhao Zhang, Mike Wilde, Ian Foster, Pete Beckman, Kamil Iskra, and Ben Clifford. Toward loosely coupled programming on petascale systems. In Proc. IEEE/ACM Supercomputing 2008, November 2008.
[3] Michael Wilde, Ian Foster, Kamil Iskra, Pete Beckman, Zhao Zhang, Allan Espinosa, Mihael Hategan, Ben Clifford, and Ioan Raicu, "Parallel scripting for applications at the petascale and beyond," *Computer*, vol. 42, pp. 50–60, 2009.

[4] Michael Wilde, Mihael Hategan, Justin M. Wozniak, Ben Clifford, Daniel S. Katz, and Ian Foster, "Swift: A language for distributed parallel scripting," *Parallel Computing*, pp. 633–652, September 2011.

[5] Karen Schuchardt, Bruce Palmer, Khushbu Agarwal, Tim Scheibe, "Many Parallel Task Computing for a Hybrid Subsurface Model," SciDAC'11 Conference, 2011.

[6] Zhiyong Zhang, Lanyuan Lu, Will G. Noid, Vinod Krishna, Jim Pfaendtner, and Gregory A. Voth, "A Systematic Methodology for Defining Coarse-Grained Sites in Large Biomolecules," *Biophysical Journal*, vol. 95, pp. 5073–5083, December 2008.

[7] Zhiyong Zhang, Jim Pfaendtner, Andrea Grafmuüller, and Gregory A. Voth, "Defining Coarse-Grained Representations of Large Biomolecules and Biomolecular Complexes from Elastic Network Models," *Biophysical Journal*, vol. 97, pp. 2327–2337, October 2009.

[8] Zhiyong Zhang and Gregory A. Voth, "Coarse-Grained Representations of Large Biomolecular Complexes from Low-Resolution Structural Data," *J. Chem. Theory Comput.*, vol. 6, pp. 2990–3002, 2010.

[9] Marius Stan, "Discovery and design of nuclear fuels," *Materials Today*, vol. 12, pp. 20-28, 2009.

[10] Justin M. Wozniak and Michael Wilde, "JETS: Language and System Support for Many-Parallel-Task Computing," Proc. Workshop on Parallel Programming Models and Systems Software for High-End Computing at ICPP, 2011.

[11] Justin Wozniak, Ketan Maheshwari, Zhao Zhang, Todd Munson, Ian Foster, Daniel S. Katz, Ewing Lusk, Matei Ripeanu, and Michael Wilde, "Parallel evaluation of dataflow programs for extreme-scale many-task computing," submitted to ACM SIGPLAN Symposium on Principles and Practice of Parallel Programming, 2012.

[12] Justin M. Wozniak and Michael Wilde, "Case Studies in Storage Access by Loosely Coupled Petascale Applications," Proc. Petascale Data Storage Workshop at SC'09, 2009.

[13] Lauro Costa and Matei Ripeanu, "Towards Automating the Configuration of a Distributed Storage System, 11th ACM/IEEE International Conference on Grid Computing (Grid 2010), October 2010.

[14] Zhao Zhang, Daniel S. Katz, Matei Ripeanu, Michael Wilde, Ian Foster, "AME: An Anyscale Many-Task Computing Engine," submitted to 6th Workshop on Workflows in Support of Large-Scale Science (WORKS11), 2011.